# Sensitivity Comparison of Macro- and Micro-electrochemical Biosensors for Human Chorionic Gonadotropin Biomarker Detection

*Samar Damiati [1,\*], Carrie Haslam [2], Sindre Søpstad [3,4], Martin Peacock [3], Toby Whitley[2], Paul Davey[2], Shakil A. Awan[2]*

[1] Department of Biochemistry, Faculty of Science, King Abdulaziz University, Jeddah, SA

[2] Wolfson Nanomaterials and Devices Laboratory, School of Computing, Electronics and Mathematics, Faculty of Science and Engineering, University of Plymouth, Plymouth, UK

[3] Zimmer and Peacock Ltd., Royston, UK

[4] Department of Microsystems, Faculty of Maritime and Natural Sciences, University of Southeast Norway, Borre, Norway

**KEYWORDS**

Electrochemical biosensors, Macro-electrodes, Micro-electrodes, Screen printed carbon electrode (SPCE), Human chorionic gonadotropin (hCG), Tumor biomarker, diagnostics.

**ABSTRACT**

Selectivity and sensitivity are important figures of merit in the design and optimization of electrochemical biosensors. The efficiency of the fabricated immunosensing surface can easily be influenced by several factors, such as detection limit, non-specific binding, and type of sensing platform. Here, we demonstrate the effect of macro- and micro-sized planner working electrodes (4 mm and 400 μm diameter, respectively) on the electrochemical behavior and the performance of the developed biosensor to detect human chorionic gonadotropin (hCG), please say why hCG here briefly. The fabricated screen-printed sensor was constructed by modifying the carbon macro- and micro-electrodes with a linker, 1-pyrenebutyric acid-N-hydroxysuccinimide ester (PANHS) and immobilization of anti-hCG antibodies to detect specifically the hCG protein. The characterization of the developed electrodes was performed by cyclic voltammetry (CV) and square wave voltammetry (SWV). Each immunesensing system has its unique electrochemical behavior which might be attributed to arrangement of particles on the surface. However, the smaller surface area of the micro-electrode is found to show higher sensitivity (1 pg/mL) compared to the macro-electrode sensor with a lower detection limit of 100 pg/mL. The proposed assay represents a promising approach that is highly effective for specific detection of an analyte and can be exploited to target biomarkers for a variety of point-of-care diagnostic applications.

## 1. INTRODUCTION

Development of selective and sensitive immunosensors to quantify a small concentration of analytes is still a challenging task due to the difficulty of controlling the fabricated sensing surface. Several factors may impact on the efficiency of the developed biosensor including the wettability properties of the sensing layer, orientation and density of the captured biomolecules (antibodies or aptamers), and the choice of the sensing platform, which must be inert to avoid any overlapping features of other materials on the surface of the sensing electrode (1-3). Electrochemical analysis is one of the most widely used techniques in clinical applications due to its speed, low cost, design flexibility, and high sensitivity and selectivity (4-7). Moreover, electrochemical-based biosensors can be exploited either to understand the electrochemical behavior of biomolecules or as effective tools to detect different analytes in clinical, environmental, agricultural, pharmaceutical and food samples (8-10). Depending on the application, several features need to be considered in the functionalized electrode such as size, shape, structure, composition, morphology and the material of the working electrode. An important parameter in developing an efficient sensor is the size of the most common planar working electrode. Performing electrochemical measurements at micron or submicron scale offer many advantages including (i) lower ohmic losses; (ii) reduced electrode capacitance due to the smaller surface area; (iii) enhanced signal-to-noise ratio; (iv) changes in the electrode surface diffusion from linear to radial which increases the mass transport rates to and from the electrode; (v) reduced polarization time which ultimately lowers the setup time and shortens the overall assay time and finally, (vi) permits electrochemical analysis of *in vivo* processes (11-13). However, despite these advantages, micro-electrodes suffer from some limitations compared to macro-electrodes such as the additional effort needed for cleaning and pretreatment (electrochemically and mechanically) to regenerate the electrode surface, their matrix materials and chemical reactions can be easily deactivated with aggressive and organic solvents and the requirement of a sensitive current amplifier for measuring currents in the micro-ampere range (14-15). They are often manufactured at the performance limits of the manufacturing process, and hence suffer a higher variability in relative electrode area compared to macro-electrodes, which directly affects their sensitivity.

Generally, when it is difficult to obtain quantitative or qualitative voltammeteric responses for the target analytes, an electro-catalytic mediator such as polymers or linkers can be used to modify the electrode surface to enhance the sensing performance of the fabricated sensor. There are several methods to immobilize mediator molecules onto the sensor including physical attachments, drop casting, physical or chemical covalent bond sorption and mixing into the carbon paste (16). 1-pyrenebutyric acid-N-hydroxy-succinimide ester (PANHS) is a linker that can be used as a scaffolding molecule to modify electrode surface by providing a suitable base for immobilizing antibodies. PANHS has been immobilized on carbon or graphene sheets (17,18), and carbon nanotubes through π-stacking and its succinimidyl ester group can react strongly to nucleophilic substitution by the amine groups of antibody or protein (19-21).

Human chorionic gonadotropin (hCG) is a 37 kDa glycoprotein hormone composed of two non-identical subunits: α-hCG (14 kDa, 92 amino acids) which can be found in different glycoprotein hormones such as thyroid stimulating hormone and β-hCG (23 kDa, 145 amino acids) which is unique to hCG. These two subunits are joined together by hydrophobic and ionic non-covalent interactions. This hormone is an important diagnostic marker of pregnancy and early fetal loss (22-24). However, β-hCG is not found in healthy men but presence of free β-subunit in blood is widely used as a tumor biomarker in selected tumors such as lung (25), stomach (26), and pancreas (27). Indeed, it has been recommended to combine blood and urine assays of β-hCG for cervical and ovarian cancers detection (28). However, hCG is expressed by both trophoblastic and non-trophoblastic human malignancies and plays a critical role in cell growth, invasion and malignancy at advanced stage of cancer (29).

Here we report on a proof-of-concept electrochemical sensor for the detection of hCG using commercially available macro- and micro-sized working electrodes (4 mm and 400 μm diameter, respectively). Screen-printed electrodes were used as they are inexpensive, disposable, and easily functionalized. Fabricated sensors and detection of hCG were characterized by cyclic voltammetry (CV) and square wave voltammetry (SWV). The developed sensor is based on carbon graphite electrodes modified with anti-hCG antibody / PANHS / SPCE (**Scheme 1**). The fabricated micro-electrode exhibits a detection limit of 1 pg/mL compared to macro-electrode with a ~100 pg/mL limit of detection (LoD). The developed assay is relatively simple, fast in response, cost effective and easy to characterize.

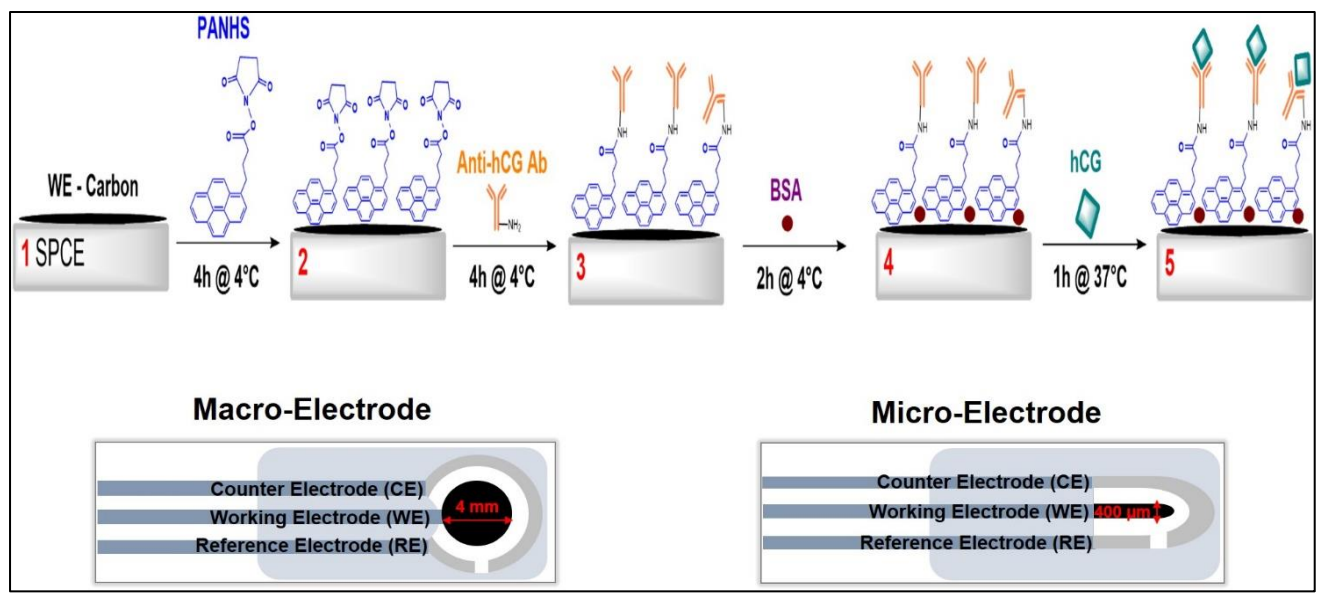

**Scheme 1:** Approach for the detection of hCG biomarker by modified macro-and micro-electrodes. Schematic showing the steps involved for the fabrication of electrochemical biosensor (not to scale) (1) bare screen printed carbon electrode (SPCE) either macro- or micro-electrode; (2) modification of electrode surface with PANHS; (3) immobilization of anti-hCG antibody; (4) blocking with BSA to prevent non-specific binding; (5) capturing of hCG antigen onto the modified surface and formation of antibody-antigen complex.

## 2. EXPERMINTAL SECTION

### 2.1. Reagents

The hCG antigen and antibody were purchased from Abcam, UK. Insulin, phosphate-buffered saline (PBS) tablets, bovine serum albumin (BSA), $K_3[Fe(CN)_6]$, $K_4[Fe(CN)_6]$, and KCl, were obtained from Sigma-Aldrich GmbH, Germany. Screen printed carbon macro- and micro-electrodes (SPCE) were obtained from DropSens (Spain) and Zimmer and Peacock (UK) respectively.

## 2.2. Preparation of the modified sensor

Three-electrode system: counter electrode (CE, carbon), reference electrode (RE, silver) and working electrode (WE, Diameter of carbon macro- and micro-electrodes = 4 mm and 400 µm, respectively) was used in this study to perform the electrochemical measurements. Initially, working electrodes were incubated with 2 mM PANHS in methanol for 4 hours at 4°C in a wet chamber. After rinsing with PBS, 20 µg/ml of anti-hCG antibodies were dropped on the surface and incubated for another 4 hours at 4°C followed by the blocking with 0.5% BSA in PBS for at least 2 hours at 4°C to minimize unspecific adsorption on the surface. A rinsing step with PBS was performed after each step. The developed sensors were stored at 4°C until further use. Different concentrations (from 1 pg/mL to 100 ng/mL) of hCG antigen in PBS were incubated with the modified electrodes for 60 min at 37°C to from antigen-antibody complex.

## 2.3. Electrochemical measurements

The sensor performance was investigated using cyclic voltammetry (CV) and square wave voltammetry (SWV). Cyclic voltammograms were recorded from -0.4 to 0.6 V for macro-electrodes and from -0.2 to 0.5 V for micro-electrodes at scan rates of 10 and 50 mV/s. Square wave voltammograms were recorded in the potential interval of 0.45 to -0.15 V under the following conditions: amplitude of 25 mV, frequency of 15 Hz, and an increase in potential of 5 mV, yielding an effective scan rate of 75 mV s$^{-1}$. All electrochemical measurements were performed in a solution of 10 mM [Fe(CN$_6$)]$^{3-}$ and 10 mM [Fe(CN$_6$)]$^{4-}$ (1:1) containing 100 mM KCl.

## 2.4. Raman spectroscopy

Raman spectra were collected using an XploRA Raman system (Horiba, Middlesex, UK) running LabSpec 6 and equipped with a 532 nm HeNe laser delivering ~4mW of laser power at the sample. An Olympus BX41 microscope (Olympus Corp., Tokyo, Japan) with a 100 µm slit width and a 300 µm confocal hole, with 1200/mm gratings and 1000x magnification

(resolution ~0.26 µm). An MPlan N 100x microscope objective (NA of 0.90 and working distance 0.21 mm) focused the laser on the sample with a spot size of ~7.5 µm diameter. A 1024×256 thermoelectrically cooled charge coupled device (CCD) camera was used for acquisition of Raman spectra at up to 1.48 MHz readout speed.

## 2.5. Contact angles measurements

The contact angles of water on both the micro- and macro-electrodes of the sensor surface were measured using a goniometer (Easy Drop, Krüss, Germany) at room temperature. Briefly, 3-5 µl of Milli-Q water was deposited onto the surface and the angle was measured immediately. All contact angle measurements were repeated in triplicate.

## 3. RESULTS AND DISCUSSION

### 3.1. Characterization of bare carbon macro- and micro-electrodes by Raman spectroscopy:

Raman spectroscopy is a highly sensitive technique that enables characterization of the molecular morphology of carbon materials and provides information about their structure and properties. In this study, Raman spectroscopy was used to analyze the carbon matrix of macro- and micro-electrodes. **Figure 1** displays the Raman spectra from the macro- and micro-electrode surfaces. Raman peaks were observed at ~1350, ~1580 $cm^{-1}$ and ~2750 $cm^{-1}$ which correspond to the disordered (D) band and graphitic (G) and 2D bands respectively, which correlate to known bands for carbon as reported earlier (30,31). The measured intensities of D, G and 2D peaks varied across different areas of the same macro or micro-electrode surfaces. In particular, the three peaks are prominent from most of the examined areas of the macro-electrode while only a few areas of the micro-electrode surface showed D peaks, which suggests the latter electrode had relatively low density of defects and disorder compared to the macro-electrode. There are several factors that can affect the Raman signal from the surface of the working electrode, such as peak position, intensity rations I(D)/I(G), I(G)/I(2D), full-width-half-maximum of the peaks, stress, doping etc.. Therefore, the nature and quality of the electrode surface is found to play a significant role in influencing the sensitivity of the biosensing surface. The Raman spectrum in Figure 1 lends support to our results from the cyclic voltammetry measurements (Section 3.6) which demonstrate that the micro-electrode is approximately an order-of-magnitude more sensitive than the macro-electrode.

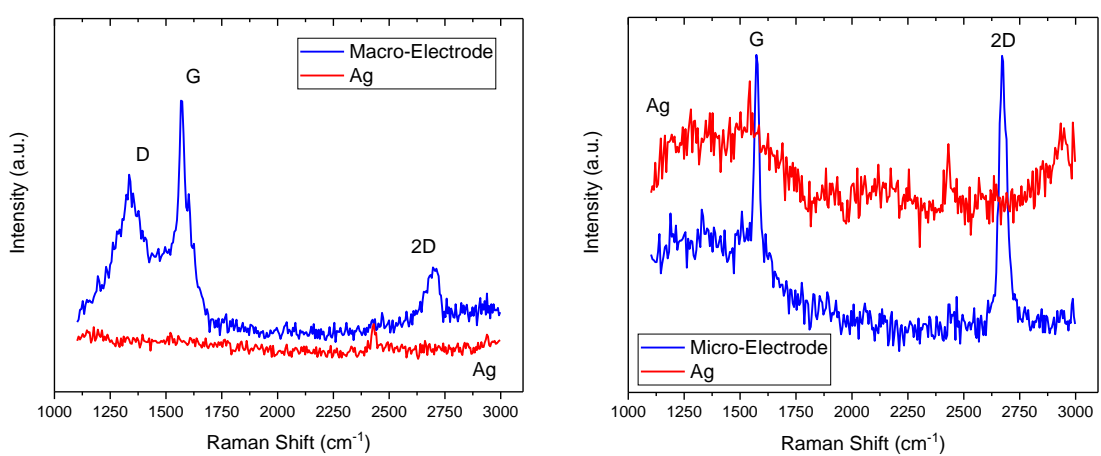

**Figure 1:** Raman spectra of bare carbon for macro- and micro-electrodes compared to Ag in locality with the working electrodes.

## 3.2. Contact angle analysis

The wettability of the modified sensor was investigated by studying the contact angles of water droplets on the carbon substrate. Generally, small angle indicates hydrophilic surface while large angle indicates hydrophobic surface. The importance of investigating the wetting properties is to determine the quality of the modified surface since the wettability has an impact on molecules immobilization (32). In this work, contact angle data for water droplet on the unmodified carbon micro-substrate showed hydrophilic surface (68.9°) which means good wettability and adhesiveness and high solid surface free energy. In contrast, unmodified carbon macro-substrate showed hydrophobic surface (97.9°). Further modification of carbon macro-surface with PANHS leads to lower hydrophilicity (87.8°) due to its structural nature. The structure of PANHS includes a pyrene and amine-reactive hydrophobic region. A stable non-covalent complex can be formed between pyrene derivatives and carbon substrate while NHS ester can be used to couple to amine-containing ligands such as antibodies. Subsequently, functionalizing the surface with hCG antibodies and blocking with BSA decreased the angles to 63.4 and 52.5°, respectively. These reduction of contact angle values indicate increasing wettability of the generated surface (**Figure 2**). The final developed layer reveals a hydrophilic matrix that is suitable for biomolecular immobilization and detection.

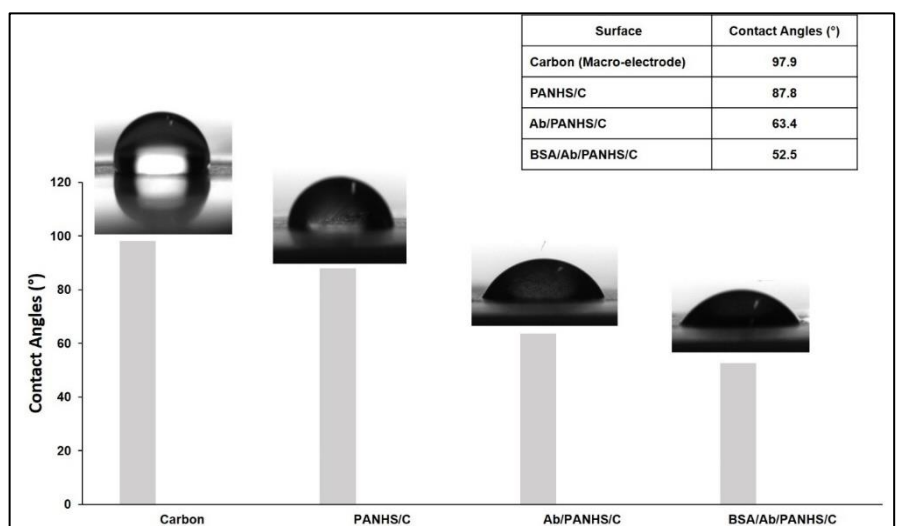

**Figure 2:** Contact angle images of water droplets on unmodified carbon macro-electrode, PANHS-modified carbon electrode, hCG antibody / PANHS-modified carbon electrode, and BSA / Ab / PANHS-modified carbon electrode.

**3.3. Characterization of bare carbon macro- and micro-electrodes by cyclic voltammetry**

Cyclic voltammetry of carbon electrodes in the presence of a redox mediator is a classic electrochemical experiment for characterizing a carbon electrode surface. In **Figure 3** there is a side by side comparison of two electrodes, a macro-carbon-electrode and a micro-carbon-electrode which would have been expected to give similarly shaped cyclic voltammograms when tested with a ferricyanide/ferrocyanide solution, but in fact the macro-electrode had a cyclic voltammogram that was distorted relative to the micro-electrode. The voltammograms of the macro-electrode attains its characteristic, symmetrical appearance from the reversible redox reaction, but with peak separations much greater than that dictated by the Nernst equation (59 mV for single electron reactions at room temperature). Please add Nernst equation here then we can mention the 59 mV for single electron reaction. This is attributed to the inherent resistance of the carbon resulting in an internal IR voltage drop in the electrode material. This contribution is typically negligible in metallic electrodes. For the micro-electrodes, there is the emergence of two sharp redox peaks. These are closer to the expected 59 mV peak separation, indicating that the resistance is significantly lowered. The reduced electrode size lowers the resistance $R$, since the distance to the conductive track is shorter, in accordance with Pouillette's law: $R = \rho \frac{L}{A}$ where $\rho$ (ohm cm) is the inherent resistivity of the carbon, L (cm) is the length of the resistive material, and A (cm$^2$) is its cross-sectional area.

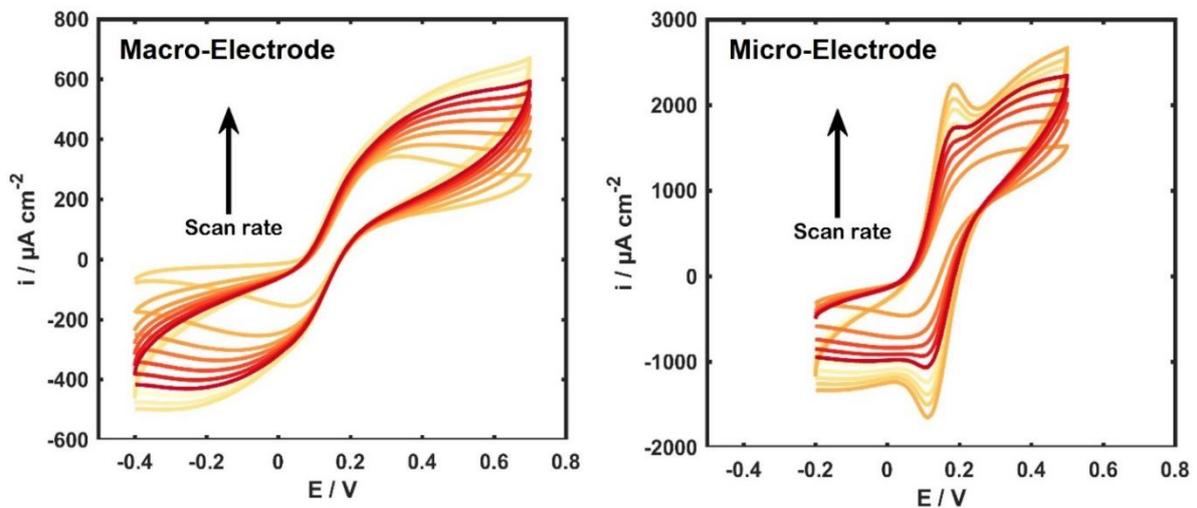

**Figure 3:** Cyclic voltammograms of the developed biosensor (C/PANHS/anti-hCG Ab/BSA) for macro- and micro-electrodes in 10 mM [Fe(CN$_6$)]$^{3-/4-}$ and 100 mM KCl with increasing scan rate from 10 to 100 mV/s.

Classical Randle equivalent circuit in **figure 4** can be used to explain the CV response data shown in Figure 3. We have added an extra resistive element, $R_{Electrode}$, to represent the resistance of the electrode. In the case of the micro-electrode the $R_{Electrode}$ element has a smaller value and does not distort the voltammograms, whilst in the case of the macro electrode the $R_{Electrode}$ value is larger and starts to dominate and distort the voltammograms by increasing the peak separation.

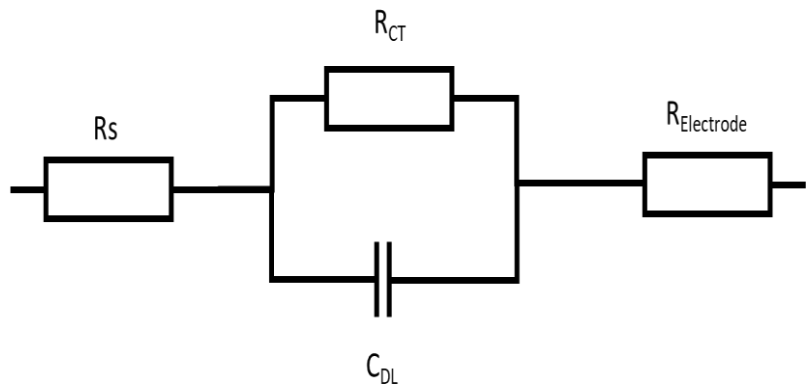

**Figure 4:** Randle Equivalent Circuit: $R_S$ solution resistance, $R_{CT}$ charge transfer resistance, $C_{DL}$ double layer capacitance, $R_{Electrode}$ resistance of the electrode.

As the macro- and the micro-electrodes are from different suppliers it is possible that the screen-printing ink is more conductive in the micro-electrode than the macro-electrode.. For either of these screen-printed electrodes to work in this application with the necessary sensitivity then it is important that the charge transfer resistance $R_{CT}$ and the double layer capacitance, $C_{DL}$, are the dominant elements in the circuit as the elements characterize the specific binding at the sensor surface during the assay, However, if the $R_{Electrode}$ value is too high relative to the $R_{CT}$ , it will mask their contributions and therefore make the assay less sensitive.

## 3.4. Sensing Assay Principles

Scheme 1 shows the current approach for the detection of hCG biomarker using PANHS-modified screen-printed carbon electrodes. Carbon micro- and macro-electrodes were modified with PANHS as a recognition layer that displayed affinity for the anti-hCG antibody and subsequently exposed to solution of the antibody. The blocking step using BSA was performed in order to minimize any unspecific binding on the surface and followed by the hCG antigen incubation step. The target hCG protein was captured by the immobilized hCG antibodies. Functionalized electrodes and detection efficiency of hCG antigen were characterized using CV and SWV measurements by monitoring Faradaic and non-faradaic currents generated by $[Fe(CN_6)]^{3-/4-}$ present in the electrolyte solution. For the macro-electrode (4 mm), **figure 5** shows the cyclic voltammograms of different components on the macro-electrode surface at scan rates of 10 and 50 mV/s. As shown in these voltammograms, typical peak-shaped current-voltage curves were obtained and an increase in the peak current of PANHS/C electrode compared to bare C electrode which can be attributed to higher electric transfer kinetics at macro-sized working electrode surface. PANHS molecules improved the electrocatalytic behavior and increased the effective surface area which enhanced electron transport between the mediator layer and electrode. By immobilizing anti-hCG antibody and blocking the electrode surface, peak current signals further reduced due to prevention of the $[Fe(CN_6)]^{3-/4-}$ redox system from approaching the electrode surface. Hence, it is clear that the intensity of anodic and cathodic peak currents is affected by each deposited layer on the sensor surface.

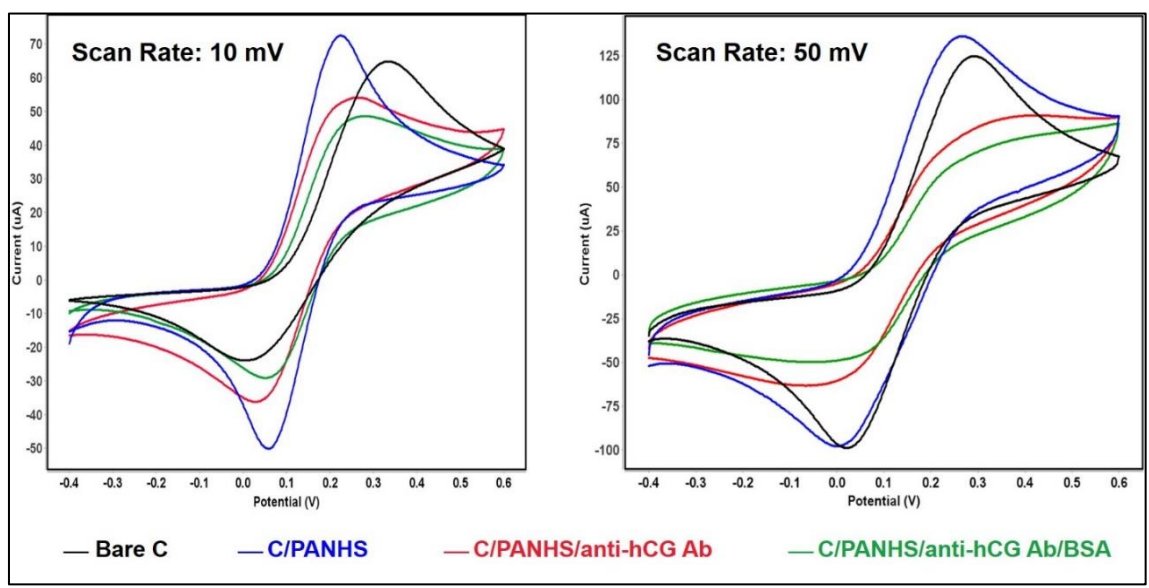

**Figure 5:** Cyclic voltammograms of macro-electrode in 10 mM $[Fe(CN_6)]^{3-/4-}$ and 100 mM KCl for each modification steps at scan rates 10 and 50 mV/s.

When the micro-electrode (400 μm) was used as a working electrode, the peak-shape is found to be significantly different to that of the macro-electrode (**Figure 6**). It is possible that the random distribution of the particles on the macro-electrode surface changes the electrode behavior while in the case of micro-electrode surface the nearest-neighbor distance minimize diffusion which leads to a higher mass transport per unit surface area. Moreover, at a lower scan rate (10 mV/s), the shape and intensity of the current-voltage curves are different compared to curves at higher scan rate (50 mV/s). However, after modifying the carbon electrode with PANHS molecules, in contrast to macro-electrode, the value of the peak current decreased which indicated formation of insulation layer that hinders the electron transfer between linker layer and the electrode. Similar to macro-electrode, the current signals decreased with the stepwise addition of the antibodies and BSA molecules. The subsequent reduction of CV signals after addition of these biomolecules indicates formation of a functional biosensor and the effect of different size of working electrodes. The results of CV studies conducted on the BSA / anti-hCG antibody / PANHS / CE biosensor obtained as a function of scan rates from 10 to 100 mV/s are shown in figure 3 for macro- and micro-electrodes.

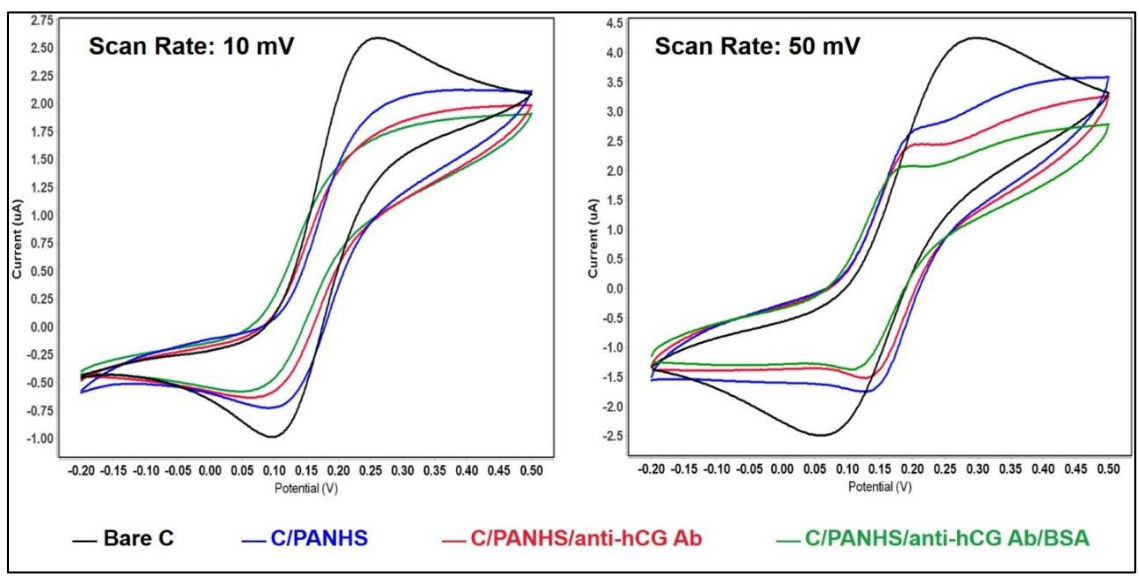

**Figure 6:** Cyclic voltammograms of micro-electrode in 10 mM $[Fe(CN_6)]^{3-/4-}$ and 100 mM KCl for each modification steps at scan rates 10 and 50 mV/s.

### 3.5. Optimization of the sensing system

Finding the optimal concentration of the biomolecules for developing sensitive and specific sensors is considered to be an important step that strongly impacts the detection limit of the sensing system. **Figure 7** shows changes of SWV current in response to detection of 10 pg/ml hCG antigen with two different concentrations of immobilized anti-hCG antibody (10 and 20 µg/ml) on the PANHS layer. In the case of macro-electrode, the SWV current reduced when higher concentration of the antibody was used (-69.25 and -81.60 µA for 10 and 20 µg/ml antibody, respectively). After addition of the hCG biomarker (10 pg/ml), the current increased (to -64.91 and -80.54 µA for 10 and 20 µg/ml antibody, respectively) which indicates that neither of these two antibody concentrations gave consistent antigen detection. In contrast, when micro-electrode was tested, both concentrations were efficient in capturing hCG protein, the current increased negatively and well-defined current peaks were obtained after hCG antigen addition. At 10 µg/ml antibody, the SWV increased from 8.50 µA to 11.88 µA after addition of 10 pg/ml hCG biomarker while the current increased from 10.43 µA to 14.50 µA at 20 µg/ml antibody concentration. Thus, higher antibody concentration, of 20 µg/ml, exhibits better efficiency to detect the target antigen. However, low antibody concentration strongly reduces the sensitivity of a designed sensor while high concentration increases the antibody layer density which may generate high steric barrier that prohibited $[Fe(CN_6)]^{3-/4-}$ ions from moving onto the electrode surface.

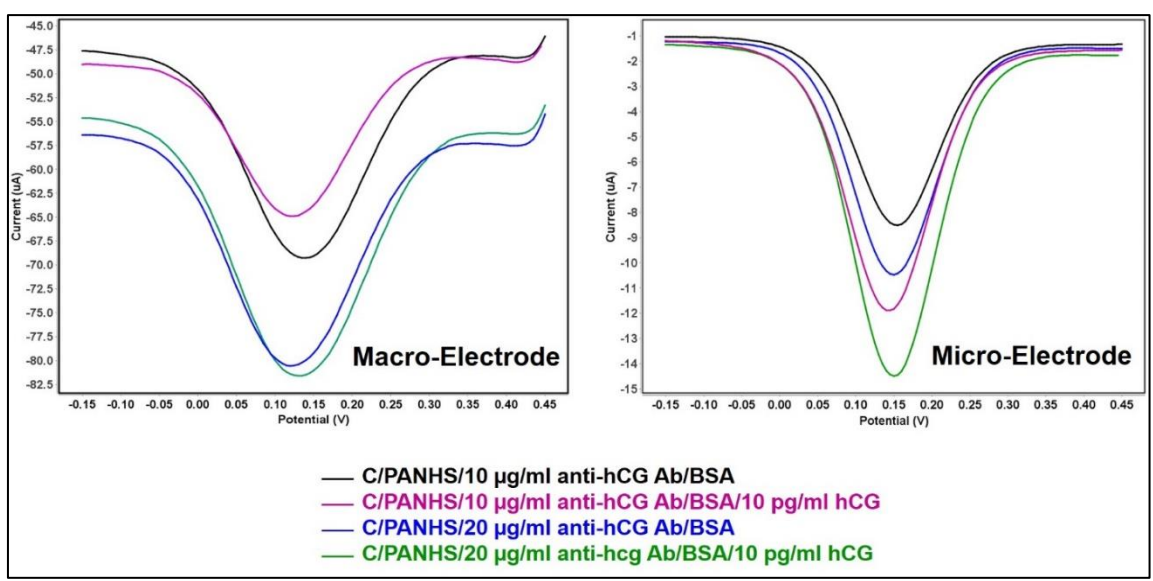

**Figure 7:** SWV signals for optimizing of the anti-hCG antibody concentrations (10 and 20 µg/ml) on the PANHS-modified macro- and micro-electrodes and detection of 10 pg/ml hCG protein in 10 mM $[Fe(CN_6)]^{3-/4-}$ and 100 mM KCl. Potential step: 5 mV; amplitude: 25 mV; frequency: 15 Hz.

### 3.6. Characterization of sensitivity of the modified electrodes

To investigate the sensitivity and analytical performance of the developed macro- and micro-biosensor, hCG biomarker was used at different concentrations in PBS and electrochemical response studies were performed in $[Fe(CN_6)]^{3-/4-}$ using SWV technique at amplification of 25 mV and frequency of 15 Hz. For macro-electrode (**Figure 8**), during the SWV measurements, the peak current reduced with increasing hCG concentrations which is attributed to formation of an antigen-antibody complex onto the developed electrode. The limit of detection was approximately 0.1 ng/ml and at higher hCG concentrations (>100 ng/mL) caused surface saturation (data not shown).

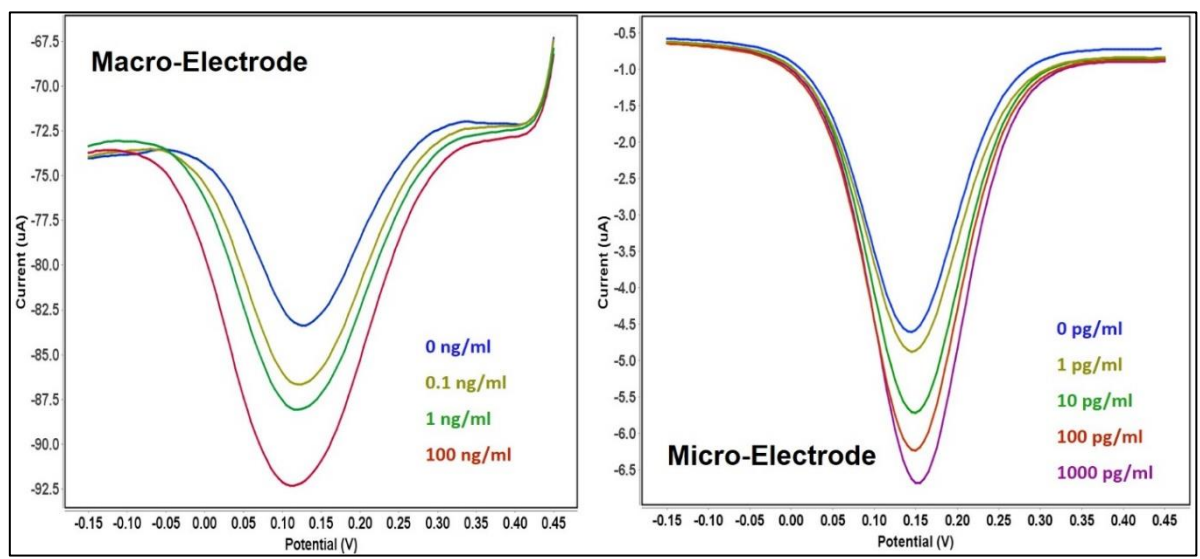

**Figure 8:** Change of SWV signals at the developed immunesensor of macro- and micro-electrodes after incubation with hCG protein at different concentrations.

Typical voltammograms are shown in figure 8 for microelectrode performance in detecting hCG at concentrations of 1, 10, 100, 1000 pg/ml. The micro-electrode shows higher sensitivity compared to the macro-electrodes. It was able to detect hCG at a concentration level as low as 1 pg/mL and sensor surface became saturated at a concentration higher than 1000 pg/ml. The obtained results confirmed the good sensitivity of the fabricated bioelectrodes. Moreover, as shown above using Raman spectroscopy, micro-electrode has better structural properties which also indicates potential for enhanced sensitivity of this type of sensing surface.

To confirm selectivity to hCG protein, the modified micro-electrode sensor was tested by incubating with 10 pg/ml insulin and its response was compared to hCG at the same concentration (**Figure 9**). The control experiment shows no significant changes after addition of insulin to the developed sensor while a significant response was recorded with hCG due to the specific interaction between the antibody and antigen.

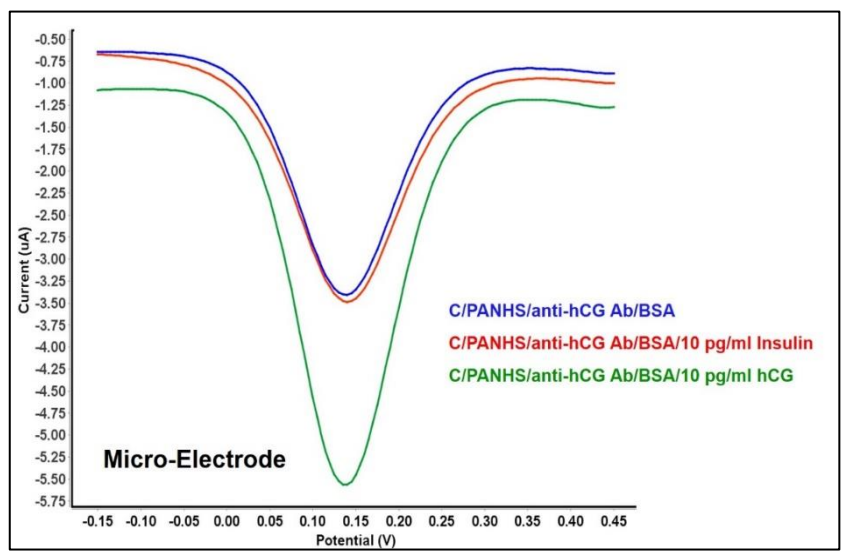

**Figure 9:** SWV signals obtained at the modified micro-electrode (C/PANHS/anti-hCG Ab/BSA) with hCG protein and insulin (as a negative control).

## CONCLUSION

The physical size of the working electrode has been demonstrated to have a direct impact on the electrochemical response and sensitivity of the fabricated biosensors using macro- and micro-electrodes. The results of CV and SWV measurements conducted on developed BSA / anti-hCG antibody / PANHS / SPCE show generation of a simple, low-cost, and label-free electrochemical biosensor with a good selectivity and sensitivity to hCG protein. The macro-electrode showed typical voltammogram peaks for the fabricated sensor and acceptable detection limit (100pg/mL), whereas the micro-electrodes showed higher sensitivity (1 pg/mL) which may be attributed to better structural properties due to the nearest-neighbour distance as confirmed by Raman spectroscopy, and the more favorable electrochemical properties of the micro-electrode relative to the macro-electrode due to the lowered electronic resistance, as evidenced by CV characterization. Future work will aim to modify the proposed sensors by using different carbon materials such as graphene and two-dimensional materials [33] to investigate the electrochemical behavior of different matrices based on the surface area, to enhance the sensitivity and to target different sample types as well as impedance spectroscopy detection techniques [34]. Indeed, the principles of the presented assay can be exploited to fabricate a variety of diagnostic sensors by using different antibodies or aptamers to recognize wide range of biomarkers.

**Supporting Information**.

# AUTHOR INFORMATION

**\* Corresponding Author:** sdamiati@kau.edu.sa

# ACKNOWLEDGMENT

Authors would like to thank Department of Nanobiotechnology (DNBT), University of Natural Resources and Life Sciences, Vienna, Austria for performing the contact angles measurements. SAA would like to acknowledge funding from the Engineering and Physical Sciences Research Council, UK (EP/M006301/1) and the University of Plymouth (GD105227).

# REFERENCES

1. Darain, F.; Gan, K.; Tjin, S. Antibody immobilization on to polystyrene substrate-on-chip immunoassay for horse IgG based on fluorescence. *Biomed Microdevices*. **2009**, 11, 653–661
2. Damiati, S.; Küpcü, S.; Peacock, M.; Eilenberger, C.; et al. Acoustic and hybrid 3D-printed electrochemical biosensors for the real-time immunodetection of liver cancer cells (HepG2). *Biosens Bioelectr*. **2017**, 94, 500–506
3. Damiati, S.; Peacock, M.; Mhana, R.; Søpstad, S.; Sleytr, U.B.; Schuster, B. Bioinspired Detection Sensor Based on Functional Nanostructures of S-Proteins to Target the Folate Receptors in Breast Cancer Cells. *Sens Actuat B*. **2018**, 267, 224-230.

4. Islam, F.; Haque, M.; Yadav, S.; Islam, M.; Gopalan, V.; Nguyen, N.; Lam, K.; Shiddiky, M. An electrochemical method for sensitive and rapid detection of FAM134B protein in colon cancer samples. *Sci Rep*. **2017**, 7, 133

5. Shiddiky, M.J., Park, H. and Shim, Y. B. Direct analysis of trace phenolics with a microchip: in-channel sample preconcentration, separation, and electrochemical detection. *Anal. Chem.* **2006,** 78, 6809–6817.

6. Lin, J.; Ju, H. Electrochemical and chemiluminescent immunosensors for tumor markers. *Biosens. Bioelectron.* **2005,** 20, 1461–1470

7. Damiati, S.; Peacock, M,; Leonhardt, S.; Baghdadi, M.A.; Damiati, L.; Becker, H.; Kodzius, R.; Schuster, B. Embedded Disposable Functionalized Electrochemical Biosensor with a 3D-Printed Flow-Cell for Detection of Hepatic Oval Cells. *Genes* **2018**, 9(2), 89.

8. Nabok, A.V.; Tsargorodskaya, A.; Hassan, A.K.; Starodub, N.F. Total internal reflection ellipsometry and SPR detection of low molecular weight environmental toxins. *Appl. Surf. Sci*. **2005**, 246, 381-386.

9. Yu, L.; Zhang, Y.; Hu, C.; Wu, H.; Yang, Y.; Huang, C.; Jia, N. Highly sensitive electrochemical impedance spectroscopy immunosensor for the detection of AFB1 in olive oil. *Food Chem*. **2015,** 176, 22.

10. Badea, M.; Floroian, L.; Restani, P.; Moga, M. Simple Surface Functionalization Strategy for Immunosensing Detection of Aflatoxin B1. *Int. J. Electrochem. Sci.* **2016,** 11, 6719- 6734

11. Scharifker, B.R. In: Bockris JO'M, Conway BE, White RE (eds) Modern aspects of electrochemistry No 22. Plenum, New York, **1992.** p 467

12. Oldham, K.B. All steady-state microelectrodes have the same "iR drop". *Electroanal. Chem.* **1987**, *237*, 303-307.

13. Bruckenstein, S. Ohmic potential drop at electrodes exhibiting steady-state diffusional currents. *Anal. Chem*. **1987**, *59*, 2098-2103.

14. Bond, A.M.; Luscombe, D.; Oldham, K.B.; Zoski, C.G. A comparison of the chronoamperometric response at inlaid and recessed disc microelectrodes. *J. Electroanal. Chem*. **1988**, *249*, 1-14.

15. Nirmaier, H.P.; Henze, G. Characteristic Behavior of Macro-, Semimicro- and Microelectrodes in Voltammetric and Chronoamperometric Measurements. *Electroanalysis* **1997**, 9

16. Metters, J.; Banks, C. Screen Printed Electrodes Open New Vistas in Sensing: Application to Medical Diagnosis. In: Schlesinger M, editor. Applications of electrochemistry in medicine, modern aspects of electrochemistry, vol. 56. New York: Springer Science+Business Media; **2013**. p. 83–120

17. Haslam, C.; Damiati, S.; Whitley, T.; Davey, P.; Ifeachor, E.; Awan S.A. Label-Free Sensors Based on Graphene Field-Effect Transistors for the Detection of Human Chorionic Gonadotropin Cancer Risk Biomarker. *Diagnostics.* **2018**, *8*, 5

18. Islam, K.; Damiati, S.; Sethi, J.; Suhail, A.; Pan, G. Development of a Label-Free Immunosensor for Clusterin Detection as an Alzheimer's Biomarker. *Sensors* **2018**, *18*(1), 308

19. Chia, J.; Tan, M.; Khiew, P.; Chin, J.; Siong, C. A bio-electrochemical sensing platform for glucose based on irreversible, non-covalent pi–pi functionalization of graphene produced via a novel, green synthesis method. *Sens Actuat B.* **2015,** 210, 558–565

20. Peng, H.; Hu, Y.; Liu, P.; Deng, Y.; Wang, P.; Chen, W.; Liu, A.; Chen, Y.; Lin, X. Label-free electrochemical DNA biosensor for rapid detection of mutidrug resistance gene based on Au nanoparticles/toluidineblue–graphene oxide nanocomposites. *Sens Actuat B.* **2015,** 207, 269–276

21. Benvidi, A.; Tezerjani, M.; Jahanbani, S.; Ardakani, M.; Moshtaghioun, S. Comparison of impedimetric detection of DNA hybridization on the various biosensors based on modified glassy carbon electrodes with PANHS and nanomaterials of RGO and MWCNTs. *Talanta.* **2016**, 147, 621–627

22. Chen, R.; Zhang, Y.; Wang, D.; Dai, H. Noncovalent sidewall functionalization of single-walled carbon nanotubes for protein immobilization, *J. Am. Chem. Soc.* **2001,** 123, 3838–3839.

23. Canfield, R.E.; O'Connor, J.F.; Birken, S.; Krichevsky, A.; Wilcox, A.J. Development of an assay for a biomarker of pregnancy and early fetal loss. Environ. Health Perspect. **1987**, 74, 57–66.

24. Cole, L.A. HCG variants, the growth factors which drive human malignancies. *Am J Cancer Res.* **2012**, 2(1), 22-35.

25. Szturmowicz, M.; Slodkowska, J.; Zych, J.; Rudzinski, P.; Sakowicz, A.; Rowinska-Zakrzewska, E. Frequency and clinical significance of beta-subunit human chorionic gonadotropin expression in non-small cell lung cancer patients. *Tumour Biol*. **1999,** 20, 99–104.

26. Rau, B.; Below, C.; Haensch, W.; Liebrich, W.; von Schilling, C.; Schlag, P.M. Significance of serum beta-hCG as a tumor marker for stomach carcinoma. *Langenbecks Arch Chir*. **1995,** 380(6), 359–364

27. Louhimo, J.; Alfthan, H.; Stenman, U.H.; Haglund, C. Serum HCG beta and CA 72-4 are stronger prognostic factors than CEA, CA 19-9 and CA 242 in pancreatic cancer. *Oncology*. **2004**, 66(2), 126–131.

28. Kinugasa, M.; Nishimura, R.; Koizumi, T.; Morisue, K.; Higashida, T.; Natazuka, T.; Nakagawa, T.; Isobe, T.; Baba, S.; Hasegaw,; K. Combination assay of urinary beta-core fragment of hCG with serum tumor markers in gynecologic cancers. *Jpn J Cancer Res.* **1995**, 86, 783–789.

29. Triozzi, P.L.; Stevens, V.C. Human chorionic gonadotropin as a target for cancer vaccines. *Oncol Rep.* **1999**, 6(1), 7–17

30. Li, J.; Tsai, C.; Kuo, S. Fabrication and characterization of inorganic silver and palladium nanostructures within hexagonal cylindrical channels of mesoporous carbon. *Polymers*. **2014**, 6(6), 1794-1809.

31. Saravanan M, Ganesan M, Ambalavanan S. An in situ generated carbon as integrated conductive additive for hierarchical negative plate of lead-acid battery. *J. power sources*. **2014,** 251, 20-29.

32. Derkus B, Emregul K, and Emregu E. Evaluation of protein immobilization capacity on various carbon nanotube embedded hydrogel biomaterials. *Mat. Sci. Engineer. C.* **2015**, 56, 132–140.

33. Awan S. A., Lombardo A., Colli., Privitera G., Kulmala T. S., Kivioja J. M., Koshino M., Ferrari A. C., Transport Conductivity of Graphene at RF and Microwave Frequencies. *2D Mater.*, **2016**, 3 (1), 015010-1-11.

34. Awan S. A., Pan G., Al Taan L. M., Li B., Jamil N., Transport electromagnetic properties of chemical vapour deposition graphene from direct current to 110 MHz. *IET Circuits, Devices Syst.*, **2015**, 9 (1), 46-51.